\documentclass[%
 aapm,
 mph,%
 amsmath,amssymb,
 reprint,%
]{revtex4-2}
\usepackage[english]{babel}
\usepackage{graphicx} 
\usepackage{epstopdf}
\usepackage{amsmath}
\usepackage{tikz}
\usepackage{hyperref}
\usepackage{subcaption}
\usepackage{float}
\usepackage{dcolumn}
\usepackage{bm}

\def\selectlanguage#1{}
\begin{document}

\preprint{APS/123-QED}

\title{Space-time excitation creates soliton trains in multimode fibers}

\author{Julien Dechanxhe}
\author{Spencer W. Jolly}%


\author{Pascal Kockaert}
 \affiliation{%
    Service OPERA-Photonique, Université libre de Bruxelles (ULB), Brussels, Belgium
 }%


\date{\today}

\begin{abstract}
In this work, we show that injecting a single space-time-coupled light pulse-beam into a multimode graded-index fiber generates a train of multimode solitons. Space-time couplings excite the spatial modes with distinct temporal profiles. Due to nonlinear interactions, with a properly chosen input power these profiles split into several unique multimode solitons. In the case of a spatially chirped input pulse, two solitons composed of modes $LP_{01}$ and $LP_{11}$ are formed. In the case of the injection of a space-time optical vortex, characterized by its topological charge $\ell$, a train composed of $|\ell|+1$ multimode solitons is generated. Their energy and modal composition are directly determined by the absolute value of the topological charge.
\end{abstract}

\maketitle

\section{Introduction}
Over the past decade, multimode optics has regained interest due to the pursuit of increased capacity of photonic networks. Since nonlinear effects are one of the main limitations to the transmission capacity, understanding multimode nonlinear processes becomes of the main interest. Beyond applications in telecommunications, nonlinear multimode propagation offers a rich avenue for creating new structures and generating collective phenomena~\cite{wright_physics_2022,krupa_multimode_2019}. Multimode nonlinear optics finds applications in a large variety of fields such as laser technology~\cite{wright_mechanisms_2020}, microscopy~\cite{cao_controlling_2023,cizmar_exploiting_2012,ploschner_seeing_2015}, and computation~\cite{tegin_scalable_2021,rahmani_learning_2022} for instance. Furthermore, multimode nonlinear interactions allow to tune supercontinuum generation~\cite{eftekhar_versatile_2017,wright_controllable_2015} and generate multimode solitons. An optical soliton is a basic structure inherent to nonlinear optics which is described by an amplitude envelope that is invariant along propagation and robust to perturbations~\cite{hasegawa_transmission_1973,song_recent_2019}. With the proper dispersion, monomode bright solitons can be generated in single-mode and multimode fibers when pumped with optical pulses. In the case of a multimode soliton~\cite{renninger_optical_2013,zitelli_optical_2024,sun_multimode_2024}, it is the combination of the modes that gives to the pulse its solitonic properties. If the energy of one of the modes forming a soliton is cut, the remaining pulse will change its temporal profile. Depending on its peak power and duration, the remaining mode can either reshape into a monomode soliton or continuously spread out temporally due to dispersion.

Generating multimode solitons requires exciting the fiber with the appropriate pulse profile in each mode, although slight variations in pulse shape are permissible due to their inherent stability. Usually, the pulse profiles are assumed to be the same or of similar shape and only varying duration. In this work, we will take advantage of the advances in the field of spatio-temporal optics to generate solitons that have drastically different temporal profiles. Spatio-temporal optics studies the effects of the non-separability between the temporal and spatial dimensions of optical fields and seeks to achieve arbitrarily non-separable structured light~\cite{shen_nonseparable_2022}. Such developments are also valuable for a wide range of applications~\cite{froula_spatiotemporal_2018,vincenti_attosecond_2012}, with a notable example being relativistic light-matter interactions~\cite{powell_relativistic_2024,piccardo_trends_2025}.
\begin{figure}[H]
    \centering
    \begin{minipage}{\textwidth}
        \begin{tikzpicture}
             \node[inner sep=0pt] (image) {
        \includegraphics[width=0.5\linewidth]{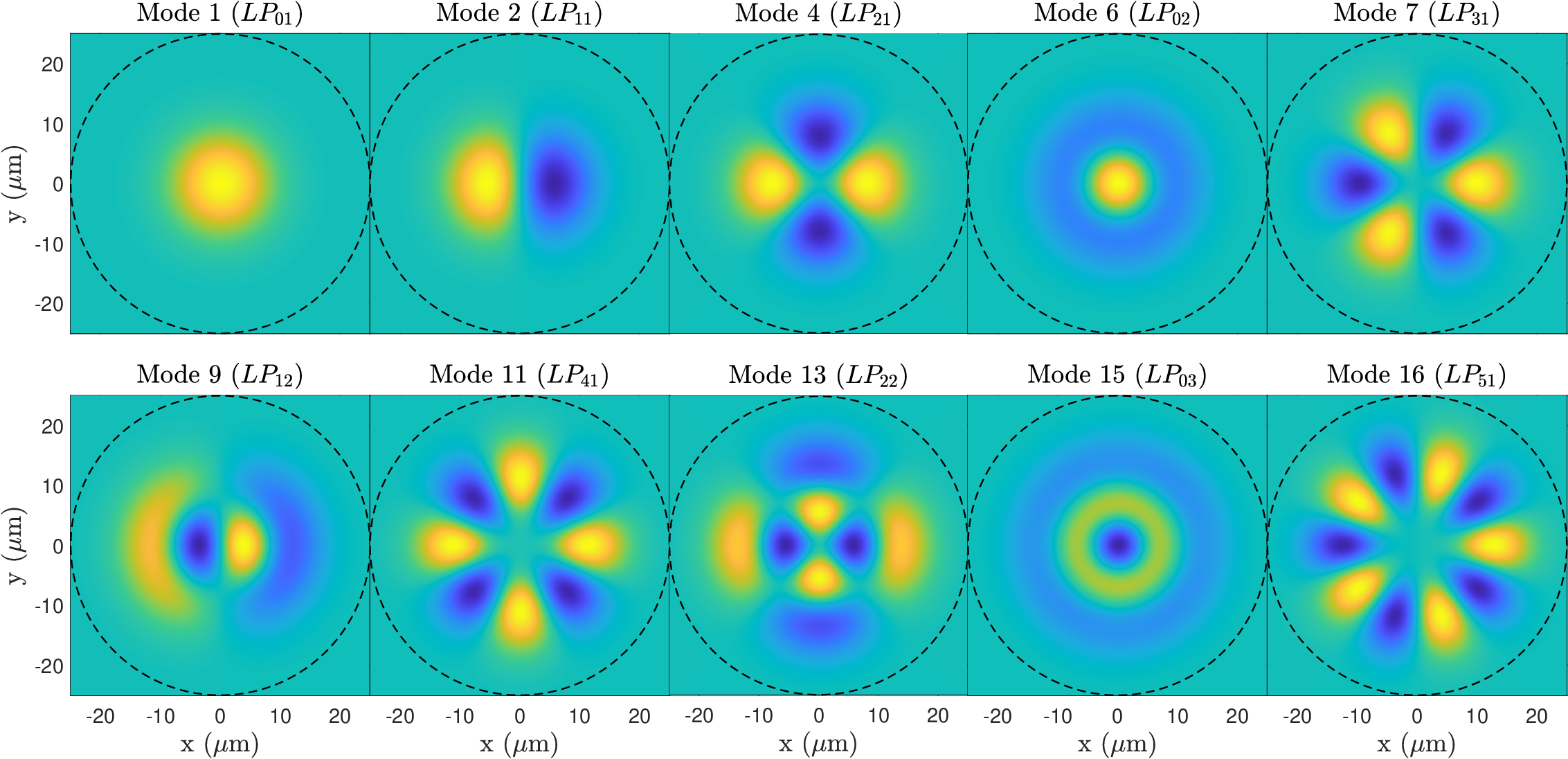}
             };
             \node[anchor=north west,yshift=10pt] at (image.north west) {a)};
        \end{tikzpicture}
        \vspace{0.3cm}
    \end{minipage}
    \begin{minipage}{0.22\textwidth}
        \begin{tikzpicture}
             \node[inner sep=0pt] (image) {
        \includegraphics[width=\linewidth]{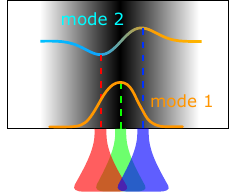}
             };
             \node[anchor=north west,xshift=-10pt] at (image.north west) {b)};
        \end{tikzpicture}
    \end{minipage}
    \begin{minipage}{0.23\textwidth}
        \begin{tikzpicture}
            \node[inner sep=0pt] (image) {
                \includegraphics[width=1.15\linewidth]{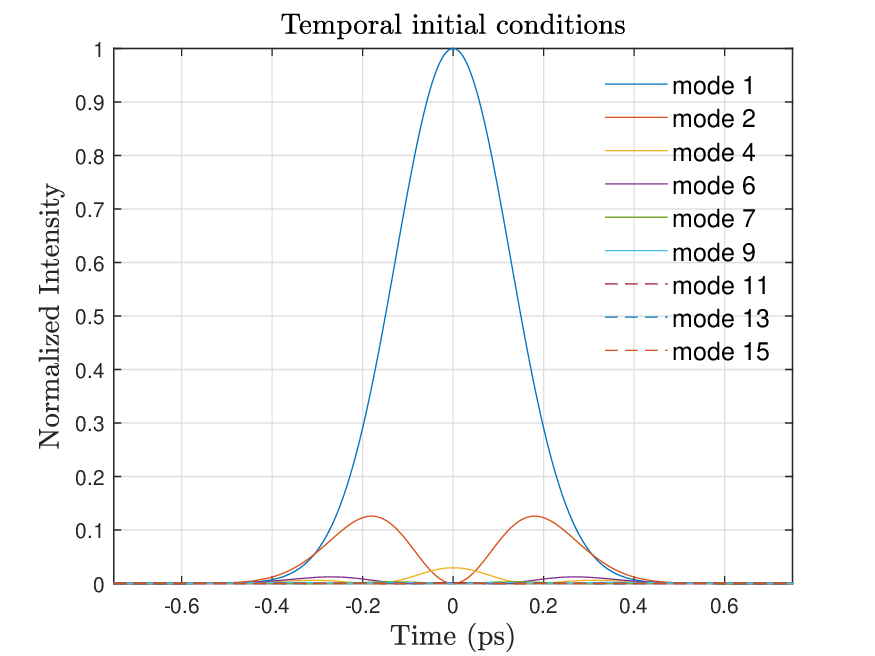}
            };
            \node[anchor= north west, xshift=-10pt,yshift=-3pt] at (image.north west) {\;\;\;\;c)};
        \end{tikzpicture}
    \end{minipage}
    \captionsetup{justification=justified, singlelinecheck=false}
    \caption{a) Transverse amplitude profile of the first 10 modes of the GRIN MMF with either cylindrical symmetry or y-axis symmetry. The black dotted line represents the core-cladding interface. The mode numbering and the corresponding LP notation are given in the title of each subplot. b) Schematic representation of the injection of a spatially chirped pulse-beam into a MMF. Each frequency overlaps differently with propagation modes. The central frequency (green) is injected at the center of the fiber. c) Normalized temporal intensity profile of the modes excited by the injection of a spatially chirped pulse with a FWHM of $250$\,fs, a waist of $7\,\mu\text{m}$, and $\tau_t/\tau_0=1$.}
    \label{Spatial_chirp_CI}
\end{figure}
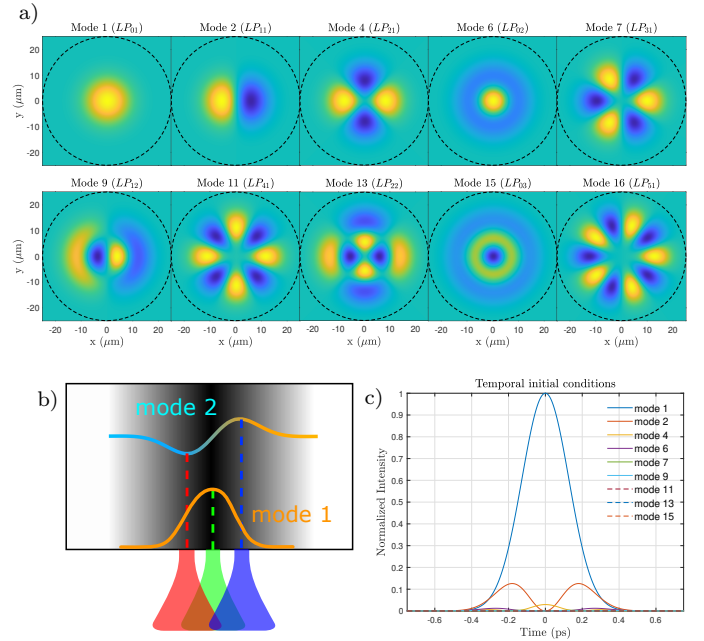
Here, we report on the generation of trains of solitons through the space-time excitation in a multimode fiber where each train contains a number of temporally spaced and unique multimode solitons. It is important to note that homogeneous optical excitation, whether single-mode or multimode, can also generate trains of solitons when the optical power becomes sufficiently large. This corresponds to the soliton number being significantly above unity. However, such a case will produce trains of identical solitons, whereby space-time excitation produces a train of multimode solitons each with different modal content.
 
To perform our study, we use the Generalized Multimode Nonlinear Schrödinger Equation (GMMNLSE) that relies on the decomposition of the input pulse into propagation modes of the fiber. All the modes are supposed to be linearly polarized in the same direction such that the problem is considered to be scalar. The model takes into account modal and chromatic dispersion, Kerr and Raman nonlinearities, and self-steepening. The Taylor expansion of the propagation constant is taken up to the fourth order. Since Raman scattering does not impact soliton train formation, the Raman strength coefficient is set to zero ($f_R=0$) in this work, which mainly changes their eventual relative propagation behavior after formation. Losses are neglected because we consider propagation over relatively short distances.

Our simulations are based on an open-source code that solves the GMMNLSE~\cite{wright_multimode_2018,wiselabaep_wiselabaepgmmnlse-solver-final_2026}. We consider a 50\,$\mu$m core diameter parabolic-graded index multimode fiber (GRIN MMF) with a 0.2 numerical aperture (NA), corresponding to a refractive index difference of $0.0137$ between the center of the core and the cladding, and a nonlinear refractive index $n_2=3.2\times10^{-20}$\,m$^2$W$^{-1}$. This fiber accepts approximately 100 propagation modes at a central wavelength of $1550$ nm, but only the first ones will be relevant here (see Fig.~\ref{Spatial_chirp_CI}.a).
\section{Spatial Chirp}
In the case of a spatially chirped input pulse, the frequency components are separated transversely across the beam profile upon focusing~\cite{jolly_coupling_2023,dechanxhe_accessing_2025}. As a result, each frequency component enters the fiber with a distinct transverse offset (see Fig.~\ref{Spatial_chirp_CI}.b). This can be induced by a set of detuned prisms before the focusing optic, for example. The spatial chirp is characterized by the pulse-front tilt ($\tau_t$) on the input beam required to produce a given spatial chirp in the focused beam, normalized to the pulse duration $\tau_0=\frac{\Delta t}{\sqrt{2\ln(2)}}$ where $\Delta t$ is the FWHM duration. In all cases, we assume $\tau_t/\tau_0 =1$. Each spatial mode is thus excited with a different spectral content leading to different temporal profiles for each mode. Fig.~\ref{Spatial_chirp_CI}.c shows normalized temporal profiles of modes excited by a spatially chirped pulse-beam with a FWHM of $250$\,fs and a waist of $7\,\mu$m at a central wavelength of $1550$\,nm. All these quantities are defined based on the case where the pulse-beam exhibits no spatial chirp—that is, when its profile is purely Gaussian. The input beam is focused at the entrance of the fiber without transverse offset, meaning that the central wavelength is injected at the center of the fiber, as shown in Fig.~\ref{Spatial_chirp_CI}.b.

As we consider space-time couplings (STC) that couple time with the transverse x-direction, the spatial modes of the fiber that are excited are either cylindrically-symmetric or symmetric with respect to the y-axis, since the input beam has a Gaussian profile along this direction. The first 10 modes exhibiting such symmetries are shown in Fig.~\ref{Spatial_chirp_CI}.a. We will refer to the modes by their numbering according to the magnitude of their propagation constant, with the fundamental mode being mode number 1. The corresponding LP notation is also given in Fig.~\ref{Spatial_chirp_CI}.a.

With spatial chirp mode 1 exhibits a Gaussian temporal profile whereas mode 2 has a first-order Hermite-Gaussian (\textit{i.e.}, having two lobes) temporal profile. These different temporal profiles will be crucial for the subsequent analysis of multimode soliton train formation. Higher-order modes, such as mode 4 and mode 6, are also excited and present different temporal profiles but their powers are too low compared to modes 1 and 2 to be impactful. As a first approach, higher-order modes will be neglected. Pulses with different FWHM duration and energy will be considered, though the shape of the temporal profiles will be similar to those shown in Fig.~\ref{Spatial_chirp_CI}.c.

\begin{figure*}[htbp]
\centering
\begin{subfigure}{0.3\textwidth}
    \centering
    \begin{tikzpicture}
        \node[inner sep=3pt] (img) {\includegraphics[width=\linewidth]{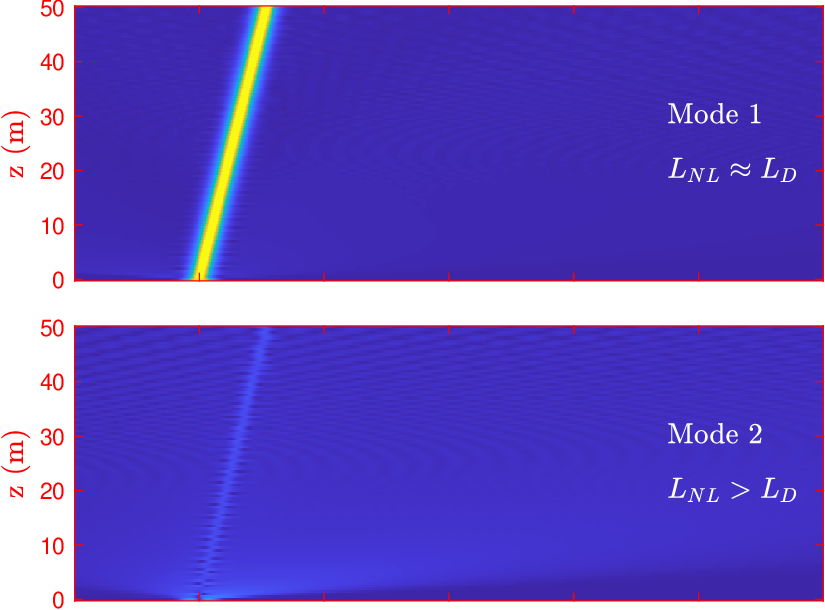}};
        \draw[black, dashed, line width=1pt] (img.south west) rectangle (img.north east);
        \node[color=white,xshift=-55pt,yshift=45pt] at (img.center) {(a)};
        \node[anchor=east,xshift=0pt] at (img.west) {\textbf{2.5 nJ}};
        \node[anchor=south] at (img.north) {\textbf{100 fs}};
    \end{tikzpicture}
\end{subfigure}
\hspace{36pt}
\begin{subfigure}{0.29\textwidth}
    \centering
    \begin{tikzpicture}
        \node[inner sep=3pt] (img) {\includegraphics[width=\linewidth]{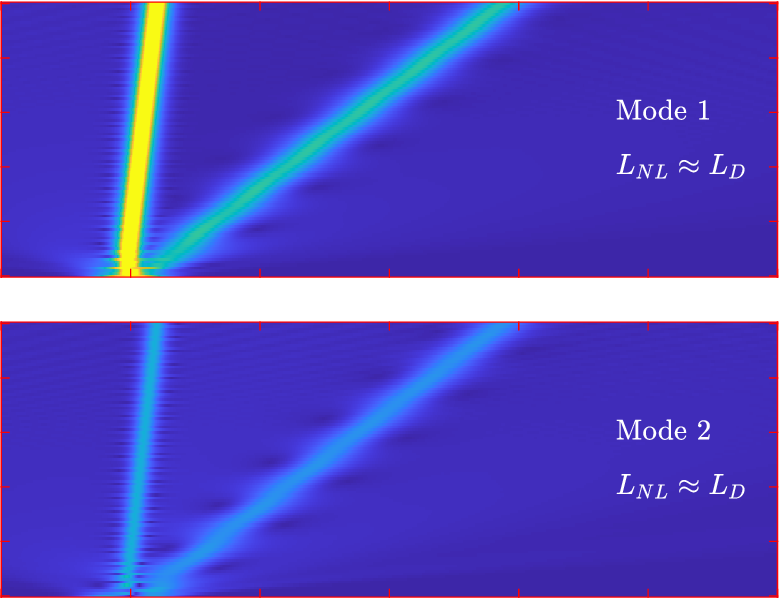}};
        \draw[black, dashed, line width=1pt] (img.south west) rectangle (img.north east);
        \node[color=white,xshift=-65pt,yshift=45pt] at (img.center) {(c)};
        \node[anchor=south] at (img.north) {\textbf{250 fs}};
    \end{tikzpicture}
\end{subfigure}
\hspace*{5pt}
\begin{subfigure}{0.29\textwidth}
    \centering
    \begin{tikzpicture}
        \node[inner sep=3pt] (img) {\includegraphics[width=\linewidth]{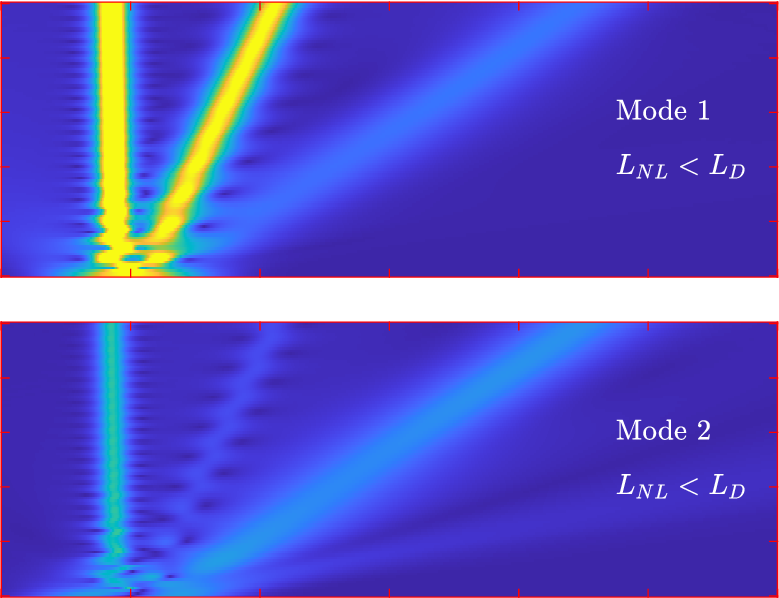}};
        \draw[black, dashed, line width=1pt] (img.south west) rectangle (img.north east);
        \node[color=white,xshift=-65pt,yshift=45pt] at (img.center) {(e)};
        \node[anchor=south] at (img.north) {\textbf{500 fs}};
    \end{tikzpicture}
\end{subfigure}
\begin{subfigure}{0.3025\textwidth}
    \centering
    \begin{tikzpicture}
        \node[inner sep=3pt] (img) {\includegraphics[width=\linewidth]{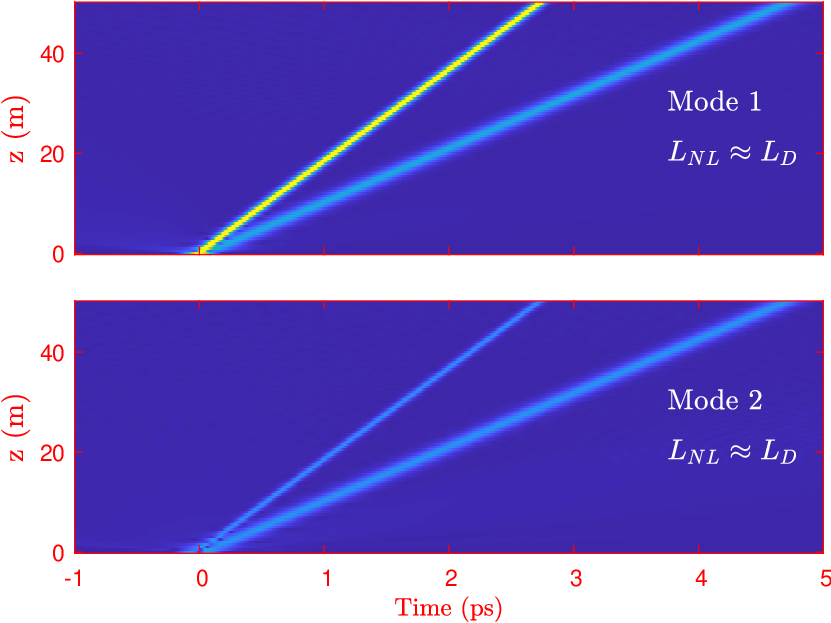}};
        \draw[black, dashed, line width=1pt] (img.south west) rectangle (img.north east);
        \node[color=white,xshift=-53pt,yshift=45pt] at (img.center) {(b)};
        \node[anchor=east,xshift=-8pt] at (img.west) {\textbf{5 nJ}};
    \end{tikzpicture}
\end{subfigure}
\hspace{36pt}
\begin{subfigure}{0.29\textwidth}
    \centering
    \begin{tikzpicture}
        \node[inner sep=3pt] (img) {\includegraphics[width=\linewidth]{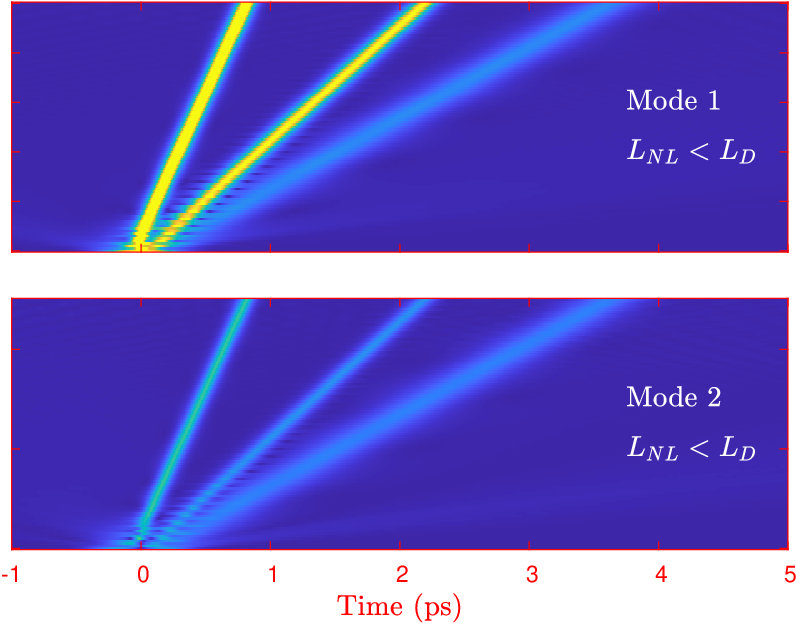}};
        \draw[black, dashed, line width=1pt] (img.south west) rectangle (img.north east);
        \node[color=white,xshift=-63pt,yshift=45pt] at (img.center) {(d)};
    \end{tikzpicture}
\end{subfigure}
\hspace*{5pt}
\begin{subfigure}{0.29\textwidth}
    \centering
    \begin{tikzpicture}
        \node[inner sep=3pt] (img) {\includegraphics[width=\linewidth]{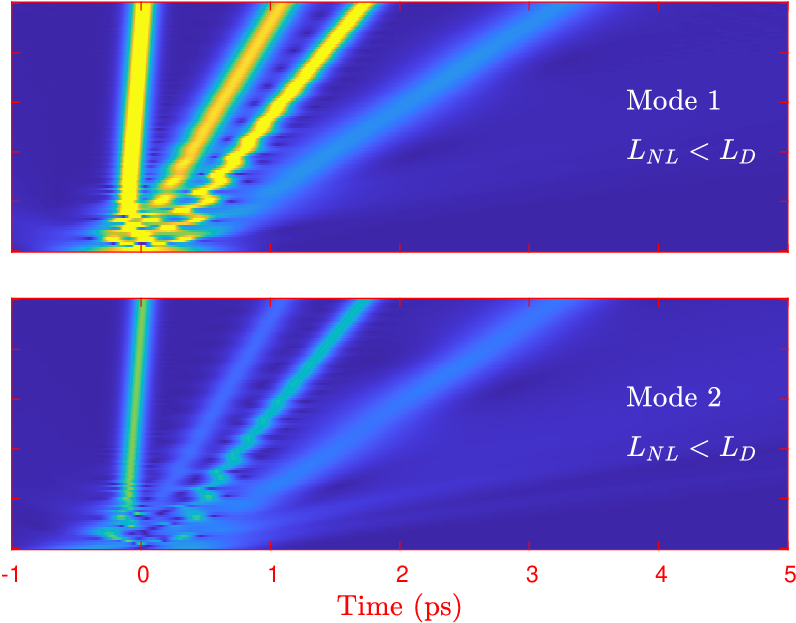}};
        \draw[black, dashed, line width=1pt] (img.south west) rectangle (img.north east);
        \node[color=white,xshift=-63pt,yshift=45pt] at (img.center) {(f)};
    \end{tikzpicture}
\end{subfigure}
\caption{Evolution of the normalized temporal intensity profiles of modes during propagation in the multimode fiber for different input pulse durations and energies. The input beam has a waist of $w_0=7\;\mu\text{m}$. The strength of the spatial chirp is the same in all cases ($\tau_t/\tau_0 =1$). a) FWHM = 100 fs and E = 2.5 nJ; b) FWHM = 100 fs and E = 5 nJ; c) FWHM = 250 fs and E = 2.5 nJ; d) FWHM = 250 fs and E = 5 nJ; e) FWHM = 500 fs and E = 2.5 nJ; f) FWHM = 500 fs and E = 5 nJ.}
\label{SC_MMS_train}
\end{figure*}
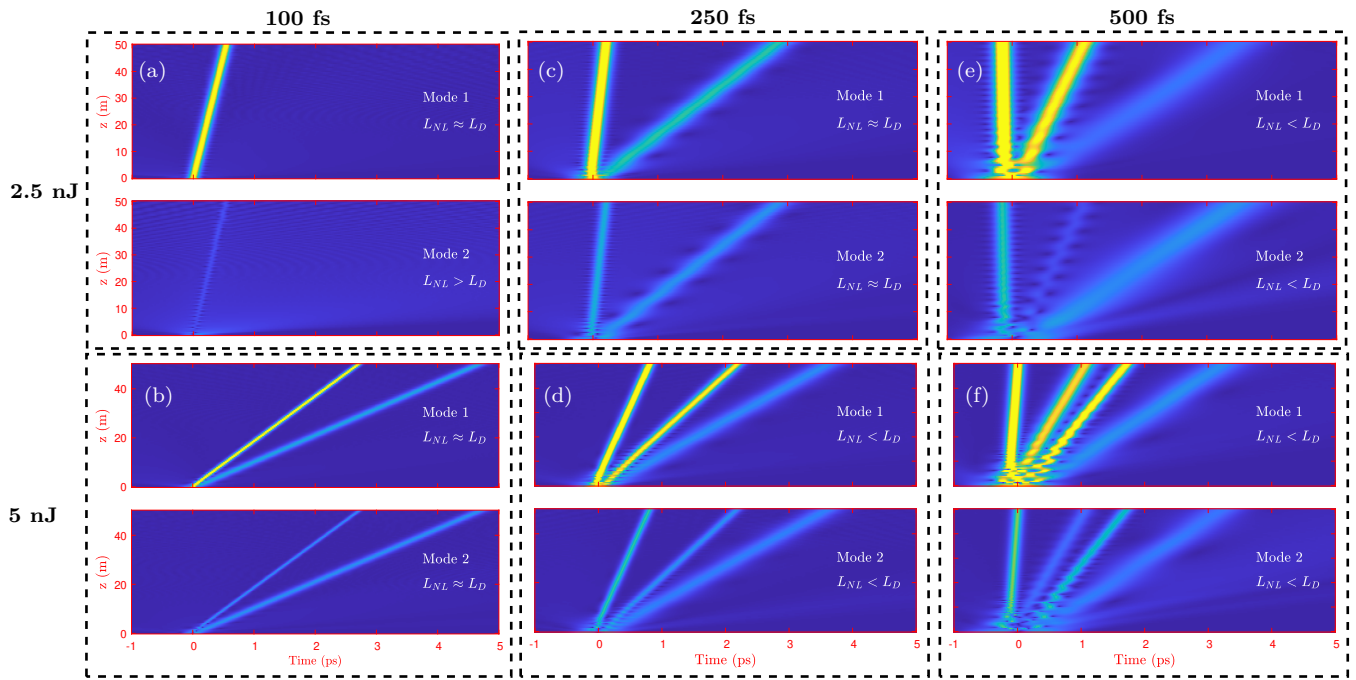

To build an understanding of the dynamics of the soliton generation in GRIN MMFs, here after, we provide an overview of the different regimes that can be observed, depending on the relative magnitude of the different characteristic lengths of the problem. A smaller length scale indicates a stronger effect. The main characteristic lengths are the dispersion length $L_{D}$ and the nonlinear length $L_{NL}$. We can also define the walk-off length $L_{W}$ and the loss length $L_{\text{loss}}$. For instance, $L_{D}< L_{NL}$ means that chromatic dispersion is dominant compared to nonlinear effects. Note that we neglect losses so $L_{D},L_{NL},L_{W}\ll L_{\text{loss}}$. Fig.~\ref{SC_MMS_train} shows the evolution of the normalized temporal intensity profiles during propagation in the multimode fiber for different input pulse durations and energies. The fully rigorous definition of these properties for multimode solitons and solitons with non-trivial temporal profiles is not yet fully developed and part of our own ongoing work, but for this purpose the basic characteristic lengths suffice.

To start with a simple case, we investigate the regime in which the group velocity dispersion (GVD) is dominant. In our simulations, this regime is obtained by using a FWHM of 100\,fs and a total energy of $2.5$\,nJ (Fig.~\ref{SC_MMS_train}.a). For the fundamental mode Kerr nonlinearity acts with the same strength as GVD ($L_{NL}\approx L_{D}$), but for mode 2 GVD is dominant ($L_{D}< L_{NL}$). Indeed, mode 1 follows a straight line in the $(t,z)$ plane while mode 2 spreads out in time. This is because mode 2 has a lower peak power than mode 1 so the nonlinearity cannot compensate the temporal spreading induced by GVD. Consequently, the fundamental mode alone forms a soliton holding a little part of mode 2 that is trapped in it due to cross-phase modulation. Hence, as could be inferred, the dominant GVD prevents the formation of multimode solitons.

Multimode solitons appear when the energy is increased to $5$\,nJ (Fig.~\ref{SC_MMS_train}.b). In such a case, mode 2 has a nonlinear length comparable to its dispersion length ($L_{NL}\approx L_{D}$). The pulse splits into two multimode solitons composed of mode 1 and mode 2. Both modes follow the same straight lines in the $(t,z)$ plane. The splitting at the very beginning of the propagation comes from the temporal profile of mode 2 that exhibits two lobes. Since we are in the anomalous dispersion regime, light tends to go where the intensity is higher. These different temporal profiles, combined with the fact that mode 2 is slower than the fundamental mode and that they interact through cross-phase modulation, result in a split of both modes into two pulses. The balance between self-phase modulation, cross-phase modulation, inter-modal dispersion, and chromatic dispersion leads to the formation of two multimode solitons with different modal compositions, peak powers, and durations. When the pulse duration is increased to $250$\,fs, for an energy of $2.5$\,nJ (Fig.~\ref{SC_MMS_train}.c), mode 2 has again a nonlinear length comparable to its dispersion length. This leads to the formation of two multimode solitons as previously explained. This time the formation comes from the fact that dispersion length is longer compared to the $100$\,fs case, hence mode 2 does not spread that much before nonlinearity acts on it. The black dots on either side of the trajectory of the slowest soliton indicate that it is not fully formed yet. Depending on their modal composition, multimode solitons can require more propagation distance before being fully formed. The transient state is characterized by the minor mode oscillating back and forth in time, bouncing against edges of the major mode, until both modes lock together. When peak powers of the modes are closer, the damping of these oscillations is weaker and the oscillation frequency is higher.

A new interesting regime is identified when the nonlinearity is further increased, with the generation of three multimode solitons, as depicted on Fig.~\ref{SC_MMS_train}.d and Fig.~\ref{SC_MMS_train}.e. These cases are closely related because the pairs of energy and duration used yield to the same ratio of dispersion length to nonlinear length. Specifically, transitioning from (d) to (e) involves halving the energy while doubling the duration. Since this ratio scales with the product of energy and duration, it remains unchanged. This relationship is not limited to exact equality: as seen in cases (b) and (c), the ratios need only to be close to each other to produce comparable effects. Finally, for a pulse duration of $500$\,fs and a total energy of $5$\,nJ (Fig.~\ref{SC_MMS_train}.f), four multimode solitons are formed. For all these three cases (d, e, and f), the initial pulse does not properly split into two distinct structures at the beginning of the propagation. When it splits into two pieces, they gather too much power compared to their equilibrium shape ($L_{NL}< L_{D}$). They emit dispersive waves and split again until there is an equilibrium between nonlinearity and dispersion leading to the formation of more multimode solitons as already depicted in~\cite{wright_spatiotemporal_2015,lu_vector_2004}. Note that the fastest multimode soliton always shows a simpler evolution than the slowest one. This asymmetry comes from inter-modal dispersion. Since mode 2 has a lower group velocity than the fundamental mode, most of its energy is used to form the slowest multimode soliton leading to a longer transient because of a richer modal composition. Note that if inter-modal dispersion was negligible compared to nonlinearities ($L_{NL} \ll L_{W}$) then modes would have split in a symmetric way.

Lastly, we address the impact of the sign of the spatial chirp on the multimode soliton train formation. Indeed, when frequency components are transversely separated, low frequencies can be either on the left of the central frequency or on its right (and vice-versa for high frequencies) depending on the sign of the pulse-front tilt $\tau_t$. In the case of an injection without a global transverse offset, the sign of the spatial chirp only impacts the spectral phase of each mode but not their temporal profile. Inverting the sign of the spatial chirp corresponds to a $\pi$-phase shift of the spectral phase of each mode. During soliton formation, modes interact through cross-phase modulation which is an incoherent coupling. Hence, the sign of the spatial chirp has no impact on the multimode soliton train formation in this case.
\section{Space-time optical vortices}
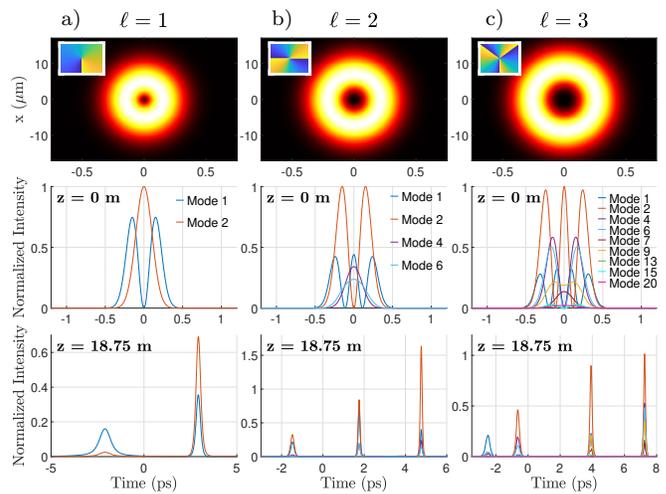
\begin{figure}[t]
    \begin{tikzpicture}
    \node[inner sep=0pt] (image){\includegraphics[width=\linewidth]{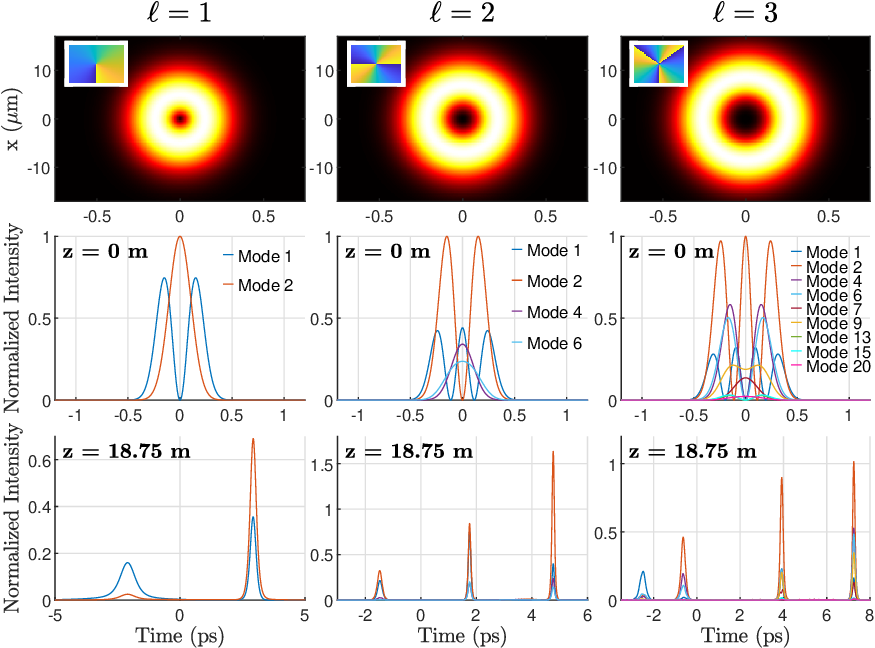}};
    \node[anchor= east,xshift = 30pt,yshift = -3.5pt] at (image.north west) {a)};
    \node[anchor= east,xshift = 110pt,yshift = -3.5pt] at (image.north west) {b)};
    \node[anchor= east,xshift = 190pt,yshift = -3.5pt] at (image.north west) {c)};
    \end{tikzpicture}
    \caption{On the first row : amplitude and phase of a space-time optical vortex for different topological charges $\ell$. Below, normalized temporal intensity profiles of propagation modes at the entrance of the fiber (second row) and after $18.75$ m of propagation (third row) for the injection of STOVs with a duration of $250$ fs. The normalization is performed against the maximum peak power at the entrance of the fiber. a) $\ell=1$ and $E=1$ nJ. b) $\ell=2$ and $E=5$ nJ. c) $\ell=3$ and $E=5$ nJ.}
    \label{figure_STOV}
\end{figure}

Space-time optical vortices (STOVs) are strongly space-time coupled structures~\cite{bekshaev_spatiotemporal_2024} that have potential applications in particle manipulation and information transfer. In the space-time plane $(t,x)$, where $x$ is an axis transverse to the propagation direction, they exhibit uniform cyclical phase that wraps around a phase singularity. As illustrated in the first row of Fig.~\ref{figure_STOV}, this phase singularity is associated to a zero-intensity region in the same space-time plane. As for spatial optical vortices, STOVs are characterized by an integer called the topological charge $\ell$, whose absolute value is equal to the number of phase jumps of $2\pi$ around the singularity and whose sign indicates the direction of the phase rotation in the space-time plane. 

Aspects of linear and nonlinear propagation of STOVs (in normal dispersion regime) in highly multimode fibers were already studied~\cite{cao_propagation_2023,zhang_optimization_2024}, including in our own work~\cite{jolly_propagation_2025}. In the anomalous dispersion regime ($\lambda_0=1550$ nm) the STOV propagation, which has not yet been studied, leads to the formation of a train of multimode solitons.

We will first discuss how a STOV of topological charge $\ell$ decomposes into different propagation modes, then we will discuss the formation of multimode soliton trains. The second row of Fig.~\ref{figure_STOV} shows modal decomposition of STOVs with different topological charges ($\ell=1,2,3$) at the entrance of the multimode fiber. The input pulse-beam has a duration is $250$\,fs and a waist of $7\,\mu\text{m}$ for all cases. As for spatial chirp, these quantities are defined for the case without STC ($\ell=0$), thus when the temporal profile is a Gaussian. The energy is 1\,nJ for the $\ell=1$ case (a) and 5\,nJ for the $\ell=2$ (b) and $\ell=3$ cases (c).

Firstly, the higher the value of the topological charge, the more modes are excited in the fiber. Secondly, for the injection of a space-time optical vortex with topological charge $\ell$, the fundamental mode temporal profile (in blue) has $|\ell|+1$ lobes. Other modes also present a multi-lobe temporal profile but with a fewer number of lobes than the fundamental mode.

Multimode solitons train emerges from the breakup of the STOV during propagation, as shown in the third row of Fig.~\ref{figure_STOV}. The splitting into several multimode soliton is due to the interplay between inter-modal dispersion and cross-phase modulation. Solitons of the same train have different group velocities and therefore separate in time during propagation. They also have each different modal compositions and peak powers. Importantly the number of multimode solitons, as well as their modal composition, depend on the absolute value of the topological charge of the injected STOV. The higher the value of $|\ell|$, the higher the number of excited modes. Since higher-order modes are the slowest ones, solitons at the end of the train are the ones containing the highest number of modes while solitons at the beginning of the train are mainly composed of lower-order modes. Slower solitons also have higher peak powers.

To understand the influence of the space-time coupling (STC) on the soliton train, we discuss in the following the STOV breakup during propagation. The train formation is dictated by the fundamental mode. Each lobe of its temporal profile gives birth to one soliton during propagation provided that the input energy is in the right range ($L_{NL}\approx L_{D}$). Since the sign of the topological charge only affects the relative phase between modes and not the shape of their temporal profiles, it does not impact the soliton train formation. A space-time optical vortex with a topological charge $\ell$ gives rise to a train of $|\ell|+1$ multimode solitons. It is important to note that the number of $|\ell|+1$ is given for a certain range of input energy for which temporal lobes from the fundamental modes have sufficient peak powers to separate each other and capture some amount of the other modes to form multimode solitons of order one. Changing the energy of the injected STOV, \textit{i.e.}~changing the peak power of the lobes, impacts the number of multimode solitons formed during propagation. If the input energy is too low, pulses forming at the beginning of the train will not have enough peak power to balance dispersion ($L_{D}<L_{NL}$) and will disperse instead of forming solitons.

On the contrary, if the input energy is too high, temporal lobes will form multimode solitons with an order higher than one ($L_{NL}<L_{D}$). In this case, temporal lobes will break into several multimode solitons and will emit dispersive waves because of excessive peak powers. This will lead to an increase of the number of multimode solitons in the train compared to the expected number of $|\ell|+1$. Furthermore, when $|\ell|$ increases the energy is distributed over more lobes, decreasing the peak power of each lobe. Indeed, increasing $|\ell|$ broadens the pulse-beam in space and time. Consequently, the higher the absolute value of the topological charge, the higher the required energy of the input space-time optical vortex to observe the formation of $|\ell|+1$ solitons.

An intriguing property of the train is that, for an energy range leading to $|\ell|+1$ solitons ($L_{NL}\approx L_{D}$), the energy of each soliton in the train is linked to the energy of the other solitons by the following rule:
 \begin{equation}\label{energy}
    E_\textrm{train} = \sum_{i=1}^{|\ell|+1} E_i=\sum_{i=1}^{|\ell|+1}iE_{1}=\frac{(|\ell|+1)(|\ell|+2)}{2}E_{1},
\end{equation}
\noindent where $E_\textrm{train}$ is the total energy of the train, $E_{i}$ is the energy of the $i^{th}$ soliton in the train (ranking from the fastest to the slowest) and $E_{1}$ is the energy of the first (fastest) soliton. In general, $E_\textrm{train}$ is not equal to the input energy since some energy remains in the form of dispersive waves or trapped radiation. Nonetheless, for the rightly chosen input energy and pulse duration, the input energy can be almost entirely transferred to the multimode soliton train. Eq.~\ref{energy} shows that when there is no STC ($\ell=0$), only one monomode soliton appears which correctly describes what happens when there is no STC. 
 
The newly identified regime comes from cases where $\ell\neq 0$. In these cases, several unique multimode solitons are generated. The energy of each soliton is directly linked to its position in the train and to the absolute value of the topological charge. The first soliton of the train has an energy $E_1$ while the second soliton has an energy $2E_1$, the third soliton has an energy $3E_1$ and so on. The last soliton of the train, which is the slowest one and with the richer modal content, has an energy $|\ell|+1$ times higher than the first one.

A main limitation of Eq.~\ref{energy} is that it holds only when $|\ell|+1$ solitons are formed. Indeed, it describes how the STC influences the energy distribution within the solitons of the train. Consequently, it provides all the information we need only when the STC is the sole factor affecting the number of solitons formed. When the structures generated at the beginning of propagation have excess energy relative to their duration ($L_{NL}<L_{D}$), they split into multiple solitons. The number of solitons in the train is then no longer solely governed by the STC, and the formula is no longer valid.
\begin{figure}[H]
    \centering
    \begin{minipage}{0.5\linewidth}
        \includegraphics[width=\linewidth]{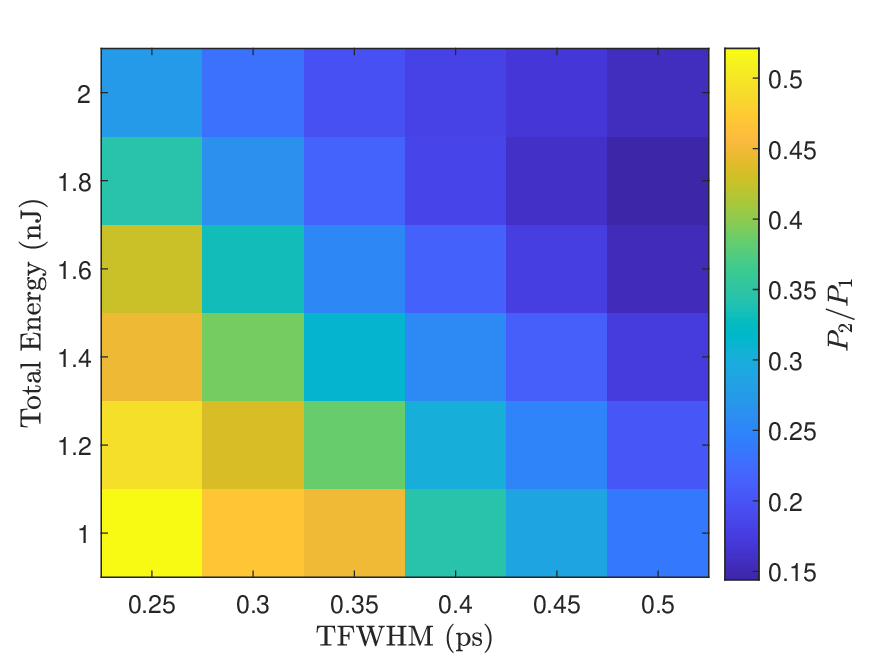}
    \end{minipage}
    \hspace*{-5pt}
    \begin{minipage}{0.5\linewidth}
        \includegraphics[width=\linewidth]{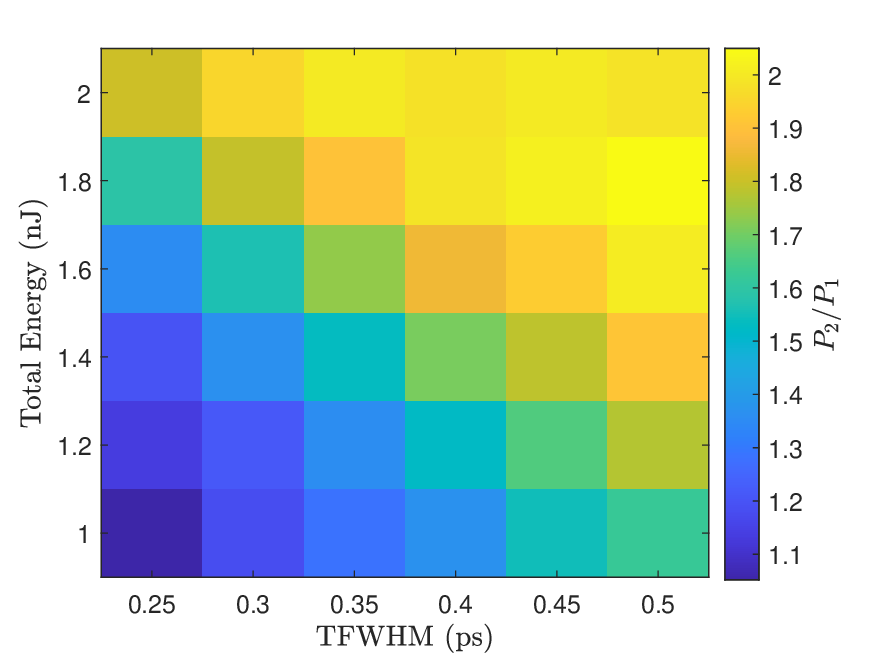}
    \end{minipage}
    \captionsetup{justification=justified, singlelinecheck=false}
    \caption{Modal power ratio of multimode solitons formed after 25 m of propagation from the injection of a STOV with topological charge $\ell=1$ for different input energies and durations. The first soliton of the train corresponds to the left plot and the second corresponds to the right plot.}
    \label{duration_STOV}
\end{figure}
Of course, each appropriate energy range is associated with a specific pulse duration since the key parameter for soliton formation is the energy-duration product. For a given input energy, it is also possible to define a duration range such that the product gives a soliton order close to one, a case that leads to the formation of $|\ell|+1$ solitons. Fig.~\ref{duration_STOV} displays the power ratio between the two constituent modes of the solitons formed by injecting STOV with a unit topological charge, for various initial pulse duration-energy pairs. For each pair, two multimode solitons are generated. At constant energy, increasing the STOV duration reduces the contribution of mode 2 in the first soliton while increasing its contribution in the second. Conversely, when the duration is held constant but the STOV energy increases, mode 2 also becomes more prominent in the second soliton. In both scenarios, the ratio of the dispersion length to the nonlinear length increases. As a result, the power ratio $P_2/P_1$ increases for the second soliton and decreases for the first. This occurs because mode 2 spreads less in time, reducing its overlap with the first lobe of mode 1. For similar duration-energy products, the power ratios in each soliton are also similar. These cases correspond to an approximately constant ratio between the dispersion length and the nonlinear length. It does also apply to spatial chirp, where the two excited modes are the same but with inverted time profile types. However, due to the significant difference in peak power (as shown in Fig.~\ref{Spatial_chirp_CI}.c), the range of duration-energy pairs leading to the formation of two first-order multimode solitons is significantly reduced.
\section{Conclusion}
In conclusion, we demonstrate that injecting a \textit{single} space-time object into a highly multimode fiber generates a train of distinct multimode solitons. We numerically studied two types of space-time couplings: spatial chirp and space-time optical vortices.

In the first case, the spatially chirped pulse-beam excites mainly modes 1 and 2 that exhibit different temporal profiles. They split into two multimode solitons during propagation for values of energy and duration that allows the formation of order one solitons.

In the second case, the injection of a space-time optical vortex with a topological charge $\ell$ leads to the generation of $|\ell|+1$ multimode solitons for peak power values that allow the generation of order one solitons. The number of solitons, their energy, and their modal composition are determined by the absolute value of the topological charge. When the soliton train arises solely from the space-time coupling, energies of solitons are ordered in descending sequence with respect to their group velocities. Thus, the fastest soliton carries the least energy, while the last soliton in the train possesses the highest energy.

In both cases, higher values of peak power induce more complex dynamics at the beginning of propagation, leading to the generation of more multimode solitons. As the sign of both space-time couplings only affects the relative phase between modes and not the shape of their temporal profiles, it does not impact the soliton train formation.

The generation of multimode soliton trains from the injection of a single space-time object is a new mechanism that could enable a high degree of control on the number of generated solitons and their properties. This work also opens new perspectives for the study of space-time couplings in multimode fibers and their impact on nonlinear dynamics.

\begin{acknowledgments}
Julien Dechanxhe is a FRIA grantee of the Fonds de la Recherche Scientifique (F.R.S.-FNRS). This work is supported by the F.R.S.-FNRS and the French Community of Belgium as part of the funding for his PhD grant.
\end{acknowledgments}

\bibliography{articles}

@article{jolly_propagation_2025,
	title = {Propagation of space-time optical vortices in multimode fibers},
	volume = {27},
	issn = {2040-8986},
	doi = {10.1088/2040-8986/adecb0},
	language = {en},
	number = {7},
	journal = {Journal of Optics},
	publisher = {IOP Publishing},
	author = {Jolly, Spencer W and Dechanxhe, Julien and Kockaert, Pascal},
	month = jul,
	year = {2025},
	pages = {075503},
}

@article{bekshaev_spatiotemporal_2024,
	title = {Spatiotemporal optical vortices: {Principles} of description and basic properties},
	volume = {9},
	issn = {2378-0967},
	shorttitle = {Spatiotemporal optical vortices},
	doi = {10.1063/5.0233758},
	number = {11},
	journal = {APL Photonics},
	author = {Bekshaev, A.},
	month = nov,
	year = {2024},
	pages = {110806},
}

@article{dechanxhe_accessing_2025,
	title = {Accessing different higher-order modes with nonlinear modal energy transfer under simple, realistic tuning of initial conditions},
	volume = {4},
	copyright = {© 2025 Optica Publishing Group},
	issn = {2770-0208},
	doi = {10.1364/OPTCON.561121},
	language = {EN},
	number = {7},
	journal = {Optics Continuum},
	publisher = {Optica Publishing Group},
	author = {Dechanxhe, Julien and Kockaert, Pascal and Jolly, Spencer W.},
	month = jul,
	year = {2025},
	pages = {1489--1504},
}

@article{krupa_multimode_2019,
	title = {Multimode nonlinear fiber optics, a spatiotemporal avenue},
	volume = {4},
	issn = {2378-0967},
	doi = {10.1063/1.5119434},
	number = {11},
	journal = {APL Photonics},
	author = {Krupa, Katarzyna and Tonello, Alessandro and Barthélémy, Alain and Mansuryan, Tigran and Couderc, Vincent and Millot, Guy and Grelu, Philippe and Modotto, Daniele and Babin, Sergey A. and Wabnitz, Stefan},
	month = nov,
	year = {2019},
	pages = {110901},
}

@article{piccardo_trends_2025,
	title = {Trends in relativistic laser–matter interaction: the promises of structured light},
	volume = {12},
	copyright = {© 2025 Optica Publishing Group},
	issn = {2334-2536},
	shorttitle = {Trends in relativistic laser–matter interaction},
	doi = {10.1364/OPTICA.558754},
	language = {EN},
	number = {6},
	journal = {Optica},
	publisher = {Optica Publishing Group},
	author = {Piccardo, Marco and Cernaianu, Mihail O. and Palastro, John P. and Arefiev, Alexey and Thaury, Cédric and Vieira, Jorge and Froula, Dustin H. and Malka, Victor},
	month = jun,
	year = {2025},
	pages = {732--752},
}

@article{jolly_coupling_2023,
	title = {Coupling to multi-mode waveguides with space-time shaped free-space pulses},
	volume = {25},
	issn = {2040-8986},
	doi = {10.1088/2040-8986/acc673},
	language = {en},
	number = {5},
	journal = {Journal of Optics},
	publisher = {IOP Publishing},
	author = {Jolly, Spencer W and Kockaert, Pascal},
	month = apr,
	year = {2023},
	pages = {054002},
}

@article{lu_vector_2004,
	title = {Vector soliton fission},
	volume = {93},
	issn = {0031-9007},
	doi = {10.1103/PhysRevLett.93.183901},
	language = {eng},
	number = {18},
	journal = {Physical Review Letters},
	author = {Lu, F. and Lin, Q. and Knox, W. H. and Agrawal, Govind P.},
	month = oct,
	year = {2004},
	pages = {183901},
}

@article{wright_physics_2022,
	title = {Physics of highly multimode nonlinear optical systems},
	volume = {18},
	copyright = {2022 Springer Nature Limited},
	issn = {1745-2481},
	doi = {10.1038/s41567-022-01691-z},
	language = {en},
	number = {9},
	journal = {Nature Physics},
	publisher = {Nature Publishing Group},
	author = {Wright, Logan G. and Wu, Fan O. and Christodoulides, Demetrios N. and Wise, Frank W.},
	month = sep,
	year = {2022},
	pages = {1018--1030},
}

@article{zitelli_optical_2024,
	title = {Optical solitons in multimode fibers: recent advances},
	volume = {41},
	copyright = {© 2024 Optica Publishing Group},
	issn = {1520-8540},
	shorttitle = {Optical solitons in multimode fibers},
	doi = {10.1364/JOSAB.528242},
	language = {EN},
	number = {8},
	journal = {JOSA B},
	publisher = {Optica Publishing Group},
	author = {Zitelli, Mario},
	month = aug,
	year = {2024},
	pages = {1655--1664},
}

@article{sun_multimode_2024,
	title = {Multimode solitons in optical fibers: a review},
	volume = {12},
	copyright = {© 2024 Chinese Laser Press},
	issn = {2327-9125},
	shorttitle = {Multimode solitons in optical fibers},
	doi = {10.1364/PRJ.531393},
	language = {EN},
	number = {11},
	journal = {Photonics Research},
	publisher = {Optica Publishing Group},
	author = {Sun, Yifan and Parra-Rivas, Pedro and Agrawal, Govind P. and Hansson, Tobias and Antonelli, Cristian and Mecozzi, Antonio and Mangini, Fabio and Wabnitz, Stefan},
	month = nov,
	year = {2024},
	pages = {2581--2632},
}

@article{renninger_optical_2013,
	title = {Optical solitons in graded-index multimode fibres},
	volume = {4},
	copyright = {2013 Springer Nature Limited},
	issn = {2041-1723},
	doi = {10.1038/ncomms2739},
	language = {en},
	number = {1},
	journal = {Nature Communications},
	publisher = {Nature Publishing Group},
	author = {Renninger, W. H. and Wise, F. W.},
	month = apr,
	year = {2013},
	pages = {1719},
}

@article{wright_spatiotemporal_2015,
	title = {Spatiotemporal dynamics of multimode optical solitons},
	volume = {23},
	copyright = {© 2015 Optical Society of America},
	issn = {1094-4087},
	doi = {10.1364/OE.23.003492},
	language = {EN},
	number = {3},
	journal = {Optics Express},
	publisher = {Optica Publishing Group},
	author = {Wright, Logan G. and Renninger, William H. and Christodoulides, Demetrios N. and Wise, Frank W.},
	month = feb,
	year = {2015},
	pages = {3492--3506},
}

@article{rahmani_learning_2022,
	title = {Learning to image and compute with multimode optical fibers},
	volume = {11},
	copyright = {De Gruyter expressly reserves the right to use all content for commercial text and data mining within the meaning of Section 44b of the German Copyright Act.},
	issn = {2192-8614},
	doi = {10.1515/nanoph-2021-0601},
	language = {en},
	number = {6},
	journal = {Nanophotonics},
	publisher = {De Gruyter},
	author = {Rahmani, Babak and Oguz, Ilker and Tegin, Ugur and Hsieh, Jih-liang and Psaltis, Demetri and Moser, Christophe},
	month = feb,
	year = {2022},
	pages = {1071--1082},
}

@article{tegin_scalable_2021,
	title = {Scalable optical learning operator},
	volume = {1},
	copyright = {2021 The Author(s), under exclusive licence to Springer Nature America, Inc.},
	issn = {2662-8457},
	doi = {10.1038/s43588-021-00112-0},
	language = {en},
	number = {8},
	journal = {Nature Computational Science},
	publisher = {Nature Publishing Group},
	author = {Teğin, Uğur and Yildirim, Mustafa and Oğuz, \{.I}lker and Moser, Christophe and Psaltis, Demetri}

@article{wright_controllable_2015,
	title = {Controllable spatiotemporal nonlinear effects in multimode fibres},
	volume = {9},
	copyright = {2015 Springer Nature Limited},
	issn = {1749-4893},
	doi = {10.1038/nphoton.2015.61},
	language = {en},
	number = {5},
	journal = {Nature Photonics},
	publisher = {Nature Publishing Group},
	author = {Wright, Logan G. and Christodoulides, Demetrios N. and Wise, Frank W.},
	month = may,
	year = {2015},
	pages = {306--310},
}

@article{eftekhar_versatile_2017,
	title = {Versatile supercontinuum generation in parabolic multimode optical fibers},
	volume = {25},
	copyright = {© 2017 Optical Society of America},
	issn = {1094-4087},
	doi = {10.1364/OE.25.009078},
	language = {EN},
	number = {8},
	journal = {Optics Express},
	publisher = {Optica Publishing Group},
	author = {Eftekhar, M. A. and Wright, L. G. and Mills, M. S. and Kolesik, M. and Correa, R. Amezcua and Wise, F. W. and Christodoulides, D. N.},
	month = apr,
	year = {2017},
	pages = {9078--9087},
}

@article{wright_mechanisms_2020,
	title = {Mechanisms of spatiotemporal mode-locking},
	volume = {16},
	copyright = {2020 The Author(s), under exclusive licence to Springer Nature Limited},
	issn = {1745-2481},
	doi = {10.1038/s41567-020-0784-1},
	language = {en},
	number = {5},
	journal = {Nature Physics},
	publisher = {Nature Publishing Group},
	author = {Wright, Logan G. and Sidorenko, Pavel and Pourbeyram, Hamed and Ziegler, Zachary M. and Isichenko, Andrei and Malomed, Boris A. and Menyuk, Curtis R. and Christodoulides, Demetrios N. and Wise, Frank W.},
	month = may,
	year = {2020},
	pages = {565--570},
}

@article{cizmar_exploiting_2012,
	title = {Exploiting multimode waveguides for pure fibre-based imaging},
	volume = {3},
	copyright = {2012 The Author(s)},
	issn = {2041-1723},
	doi = {10.1038/ncomms2024},
	language = {en},
	number = {1},
	journal = {Nature Communications},
	publisher = {Nature Publishing Group},
	author = {Čižmár, Tomáš and Dholakia, Kishan},
	month = aug,
	year = {2012},
	pages = {1027},
}

@article{ploschner_seeing_2015,
	title = {Seeing through chaos in multimode fibres},
	volume = {9},
	copyright = {2015 Springer Nature Limited},
	issn = {1749-4893},
	doi = {10.1038/nphoton.2015.112},
	language = {en},
	number = {8},
	journal = {Nature Photonics},
	publisher = {Nature Publishing Group},
	author = {Plöschner, Martin and Tyc, Tomáš and Čižmár, Tomáš},
	month = aug,
	year = {2015},
	pages = {529--535},
}

@article{cao_controlling_2023,
	title = {Controlling light propagation in multimode fibers for imaging, spectroscopy, and beyond},
	volume = {15},
	copyright = {© 2023 Optica Publishing Group},
	issn = {1943-8206},
	doi = {10.1364/AOP.484298},
	language = {EN},
	number = {2},
	journal = {Advances in Optics and Photonics},
	publisher = {Optica Publishing Group},
	author = {Cao, Hui and Čižmár, Tomáš and Turtaev, Sergey and Tyc, Tomáš and Rotter, Stefan},
	month = jun,
	year = {2023},
	pages = {524--612},
}

@article{shen_nonseparable_2022,
	title = {Nonseparable {States} of {Light}: {From} {Quantum} to {Classical}},
	volume = {16},
	copyright = {© 2022 The Authors. Laser \& Photonics Reviews published by Wiley-VCH GmbH},
	issn = {1863-8899},
	shorttitle = {Nonseparable {States} of {Light}},
	doi = {10.1002/lpor.202100533},
	language = {en},
	number = {7},
	journal = {Laser \& Photonics Reviews},
	author = {Shen, Yijie and Rosales-Guzmán, Carmelo},
	year = {2022},
	pages = {2100533},
}

@article{vincenti_attosecond_2012,
	title = {Attosecond {Lighthouses}: {How} {To} {Use} {Spatiotemporally} {Coupled} {Light} {Fields} {To} {Generate} {Isolated} {Attosecond} {Pulses}},
	volume = {108},
	shorttitle = {Attosecond {Lighthouses}},
	doi = {10.1103/PhysRevLett.108.113904},
	number = {11},
	journal = {Physical Review Letters},
	publisher = {American Physical Society},
	author = {Vincenti, H. and Quéré, F.},
	month = mar,
	year = {2012},
	pages = {113904},
}

@article{froula_spatiotemporal_2018,
	title = {Spatiotemporal control of laser intensity},
	volume = {12},
	copyright = {2018 The Author(s)},
	issn = {1749-4893},
	doi = {10.1038/s41566-018-0121-8},
	language = {en},
	number = {5},
	journal = {Nature Photonics},
	publisher = {Nature Publishing Group},
	author = {Froula, Dustin H. and Turnbull, David and Davies, Andrew S. and Kessler, Terrance J. and Haberberger, Dan and Palastro, John P. and Bahk, Seung-Whan and Begishev, Ildar A. and Boni, Robert and Bucht, Sara and Katz, Joseph and Shaw, Jessica L.},
	month = may,
	year = {2018},
	pages = {262--265},
}

@article{powell_relativistic_2024,
	title = {Relativistic {Electrons} from {Vacuum} {Laser} {Acceleration} {Using} {Tightly} {Focused} {Radially} {Polarized} {Beams}},
	volume = {133},
	doi = {10.1103/PhysRevLett.133.155001},
	number = {15},
	journal = {Physical Review Letters},
	publisher = {American Physical Society},
	author = {Powell, Jeffrey and Jolly, Spencer W. and Vallières, Simon and Fillion-Gourdeau, François and Payeur, Stéphane and Fourmaux, Sylvain and Lytova, Marianna and Piché, Michel and Ibrahim, Heide and MacLean, Steve and Légaré, François},
	month = oct,
	year = {2024},
	pages = {155001},
}

@article{wright_multimode_2018,
	title = {Multimode {Nonlinear} {Fiber} {Optics}: {Massively} {Parallel} {Numerical} {Solver}, {Tutorial}, and {Outlook}},
	volume = {24},
	issn = {1558-4542},
	shorttitle = {Multimode {Nonlinear} {Fiber} {Optics}},
	doi = {10.1109/JSTQE.2017.2779749},
	number = {3},
	journal = {IEEE Journal of Selected Topics in Quantum Electronics},
	author = {Wright, Logan G. and Ziegler, Zachary M. and Lushnikov, Pavel M. and Zhu, Zimu and Eftekhar, M. Amin and Christodoulides, Demetrios N. and Wise, Frank W.},
	month = may,
	year = {2018},
	pages = {1--16},
}

@article{hasegawa_transmission_1973,
	title = {Transmission of stationary nonlinear optical pulses in dispersive dielectric fibers. {I}. {Anomalous} dispersion},
	volume = {23},
	issn = {0003-6951, 1077-3118},
	doi = {10.1063/1.1654836},
	language = {en},
	number = {3},
	urldate = {2026-03-02},
	journal = {Applied Physics Letters},
	author = {Hasegawa, Akira and Tappert, Frederick},
	month = aug,
	year = {1973},
	pages = {142--144},
}

@article{song_recent_2019,
	title = {Recent progress of study on optical solitons in fiber lasers},
	volume = {6},
	issn = {1931-9401},
	doi = {10.1063/1.5091811},
	number = {2},
	urldate = {2026-03-02},
	journal = {Applied Physics Reviews},
	author = {Song, Yufeng and Shi, Xujie and Wu, Chengfa and Tang, Dingyuan and Zhang, Han},
	month = may,
	year = {2019},
	pages = {021313},
}

@misc{wiselabaep_wiselabaepgmmnlse-solver-final_2026,
	title = {{GMMNLSE}},
	url = {https://github.com/WiseLabAEP/GMMNLSE-Solver-FINAL},
	note = {{h}ttps://github.com/WiseLabAEP/GMMNLSE-Solver-FINAL},
}

@article{cao_propagation_2023,
	title = {Propagation of transverse photonic orbital angular momentum through few-mode fiber},
	volume = {5},
	issn = {2577-5421, 2577-5421},
	url = {https://www.spiedigitallibrary.org/journals/advanced-photonics/volume-5/issue-3/036002/Propagation-of-transverse-photonic-orbital-angular-momentum-through-few-mode/10.1117/1.AP.5.3.036002.full},
	doi = {10.1117/1.AP.5.3.036002},
	number = {3},
	journal = {Advanced Photonics},
	publisher = {SPIE},
	author = {Cao, Qian and Chen, Zhuo and Zhang, Chong and Chong, Andy C. and Zhan, Qiwen},
	month = apr,
	year = {2023},
	pages = {036002},
}

@article{zhang_optimization_2024,
	title = {Optimization of {Transverse} {OAM} {Transmission} through {Few}-{Mode} {Fiber}},
	volume = {11},
	copyright = {http://creativecommons.org/licenses/by/3.0/},
	issn = {2304-6732},
	url = {https://www.mdpi.com/2304-6732/11/4/328},
	doi = {10.3390/photonics11040328},
	language = {en},
	number = {4},
	journal = {Photonics},
	publisher = {Multidisciplinary Digital Publishing Institute},
	author = {Zhang, Chong and Cao, Qian and Zhan, Qiwen},
	month = apr,
	year = {2024},
	pages = {328},
}

\end{document}